# Optical pumping of electronic and nuclear spin in single charge-tunable quantum dots.

A. S. Bracker[1], E. A. Stinaff[1], D. Gammon[1], M. E. Ware[1], J. G. Tischler[1], A. Shabaev[1], Al. L. Efros[1], D. Park[1], D. Gershoni[2], V. L. Korenev[3], I. A. Merkulov[3]

[1]*Naval Research Laboratory, Washington, DC 20375*, USA

[2] *Physics Department, Technion-Israel Institute of Technology, Haifa, 32000, Israel*

[3] *A. F. Ioffe Physico-Technical Institute, St. Petersburg, 194021, Russia*

Abstract

We present a comprehensive examination of optical pumping of spins in individual GaAs quantum dots as we change the charge from positive to neutral to negative using a Schottky diode. We observe that photoluminescence polarization memory has the same sign as the net charge of the dot. Optical pumping of ground state electron spins enhances this effect, as demonstrated through the first measurements of the Hanle effect on an individual quantum dot. With the Overhauser effect in a high longitudinal magnetic field, we demonstrate efficient optical pumping of the quantum dot's nuclear spins for all three charge states.

Semiconductor spins have been studied by means of optical spectroscopy with remarkable success for over three decades [1]. Recently, this research field has been energized by the vision of harnessing spin in nanostructures for quantum information processing. In particular, individual carrier spins in semiconductor quantum dots (QDs) can be addressed as qubits and manipulated through optically excited states (charged excitons) [2,3,4,5,6]. In these methods, qubit initialization and readout of an electron spin can be achieved with the classic techniques of optical orientation [1].

In a direct-gap semiconductor, optical orientation begins with the conversion of circularly polarized light into a spin-polarized electron-hole pair. This polarization may be reduced by spin relaxation processes, with the remaining excitonic polarization imprinted on the polarization of luminescence. Importantly however, spin relaxation can transfer polarization to other spin degrees of freedom such as ground state electron or nuclear spins. These optically pumped spin polarizations may persist long after excitonic recombination is over.

The techniques of optical orientation have been widely applied to semiconductor bulk and quantum well systems [1], but there are relatively few reports of such work in QDs. These include direct measurements of excitonic polarization in charged QD ensembles [7,8,9,10,11,12] and single QDs [13], and demonstrations of optical pumping of electrons [7,8] and nuclei [14]. In this work, we make the connection between excitonic, electron, and nuclear polarization, all within a single QD embedded in a diode structure. We begin with a discussion of the dramatic changes in excitonic polarization that occur as positive or negative charge is injected into a QD. We then use measurements of the Hanle and Overhauser effects to demonstrate optical pumping of electron and nuclear spins, respectively.

The QDs that we investigate are defined by monolayer-high steps at the interfaces of a 3 nm GaAs/AlGaAs quantum well [15]. The quantum well layer is contained within a Schottky diode heterostructure that provides control of charge injection with an applied bias [16]. Samples were excited with circularly polarized light from a Ti:sapphire laser tuned into the quasi-continuum above the lateral QD potential barrier. The PL polarization (or "polarization memory") [1,7] is $\rho = (I_+ - I_-)/(I_+ + I_-)$, where $I_+(I_-)$ is the PL intensity measured after passing through a right (left) circular polarization analyzer. The light is then dispersed in a spectrometer and detected with a multichannel CCD, except for the measurement of the Hanle effect, where a photon counting avalanche photodiode was phase-locked to a 40 KHz polarization modulation of the exciting light to prevent nuclear polarization.

Positive circularly polarized laser light ($\sigma^+$) at normal incidence produces spin "up" holes h($\Uparrow$), with total angular momentum projection $m_h = +3/2$ along the growth direction, and spin "down" electrons e($\downarrow$), with $m_e = -1/2$. An electron-hole pair, or exciton, is captured into the charged or neutral QD and recombines there, emitting characteristic polarized photoluminescence (PL). The charged excitons $X^+$ and $X^-$ (often called trions because they consist of three particles) are shown schematically in Fig. 1(c), together with the neutral exciton $X^o$. The lowest energy negative (positive) trion consists of two paired electrons (holes) in a singlet configuration and an unpaired hole (electron). The PL polarization of $X^-$ is determined by the spin of the hole, while that of $X^+$ is determined by the electron.

The trion states of an individual QD are clearly identified by the energy of the emitted photon [15,16,17,18,19,20,21,22,23], which is shifted relative to the emission from a neutral exciton. The PL spectrum in Fig. 1(a) shows the positive trion ($X^+$), neutral exciton ($X^o$), and negative trion ($X^-$) over different bias ranges. The PL polarization changes sign when the QD charge changes sign, even though the $\sigma^+$-polarized laser always excites e($\downarrow$)-h($\Uparrow$) pairs. Around 4V bias, the polarization [Fig. 1(b)] is positive for $X^+$, roughly zero for $X^o$, and negative [7,8,10] for $X^-$. The $X^-$ shows the richest behavior. At the highest laser intensity (open circles), the $X^-$ polarization is negative for all values of the bias. At lower laser intensity (solid circles), the polarization is negative only near the charging threshold, but changes to positive with applied bias as unpolarized electrons are injected. Finally, for lower laser excitation energy (15 meV above the neutral exciton line instead of 28 meV above), the $X^-$ polarization is positive for all biases (dotted line).

Negative polarization of $X^-$ implies that the heavy-hole spin has flipped prior to recombination, while positive polarization of $X^+$ implies that the electron spin has not flipped. The heavy-hole spin flip time is much shorter than that of the electron due to spin-orbit interaction in the valence band states. This is a well-established behavior of 2D charge carriers with excess energy (hot carriers) [24]. In addition, however, negative polarization of $X^-$ actually implies that a spin-flipped hole h($\Downarrow$) contributes to $X^-$ formation with higher probability than a non-flipped hole h($\Uparrow$). Previous work has indeed shown that non-radiative dark excitons $X^o(\Downarrow\downarrow)$ can play the role of accumulating spin-flipped holes [7]. Essentially, photogenerated bright excitons $X^o(\Uparrow\downarrow)$ recombine radiatively within a short time, while dark excitons $X^o(\Downarrow\downarrow)$ arising from a hot-hole spin flip survive much longer and have a greater chance of being trapped in a charged QD. The magnitude and even the sign of X⁻ polarization depend on the relative probability for a charged QD to capture a bright or dark exciton. For example, at high bias we electrically inject many unpolarized electrons, which reduces the availability of dark excitons for capture into the QD, because they can easily recombine by forming the trions X⁻($\Downarrow\downarrow\uparrow$) elsewhere in the QW [7]. The X⁻ polarization in the QD therefore becomes positive [Fig. 1(b), solid circles]. Lower excitation energy inhibits the hole spin flip that produces dark excitons in the first place, also making X⁻ polarization more positive [Fig. 1(b) (dotted line)].

Notably, the capture of a dark exciton X⁰($\Downarrow\downarrow$) by the QD changes the spin state of the QD electron: the formation of singlet X⁻($\Downarrow\downarrow\uparrow$) requires the presence of a QD electron e($\uparrow$), but after X⁻ recombination, the spin down electron e($\downarrow$) is left. This corresponds, in the language of atomic physics, to *optical pumping* of the ground state electron spin.

With the Hanle effect, we directly measure optical pumping of the electron spin. A small transverse magnetic field ($B_x$, in the QD plane) erases the contribution of the oriented electron to the PL polarization [1,7,11]. Paired carriers are not affected by the magnetic field because their net spin is zero, and unpaired holes are not affected because their g-factor is nearly zero in the plane of the QD [15]. Depolarization of luminescence occurs when $B_x$ is large enough to make the electron spin precess faster than the rate of intrinsic dephasing and recombination. In the simplest case, the depolarization curve is Lorentzian with halfwidth $B_{1/2} = \hbar / T_s g_e \mu_B$, where $T_s$ is the electron spin lifetime, $g_e$ is its g-factor (~0.2) and $\mu_B$ is the Bohr magneton (59 μeV/T).

There is a marked difference between the Hanle linewidths for $X^-$ and $X^+$ in a single QD [Fig. 2(a)]. For $X^+$, an unpaired electron is present in the trion itself, and its lifetime is limited by fast radiative recombination. The broad Hanle feature (3.5 kG halfwidth) corresponds to a lifetime of 150 ps, roughly consistent with the expected radiative recombination time for these large GaAs QDs [25]. In contrast, a typical Hanle peak for $X^-$ is very sharp. At low laser powers we measure a halfwidth of 35 G, corresponding to a lifetime of 16 ns. This lifetime is too long to be associated with recombination and indicates that the Hanle effect depolarizes the unpaired ground state electron, which influences subsequent $X^-$ formation [26]. The $X^-$ itself does not respond to the transverse field because it has no unpaired electrons.

The lifetime obtained from the $X^-$ Hanle effect can be compared to what we expect for a localized electron influenced by fluctuations in the local nuclear spin environment [27,28]. The nuclear spins are static during an optical cycle but fluctuate during the long measurement time. The fluctuating spins, via the hyperfine interaction, behave like an effective magnetic field and lead to spectral diffusion that broadens the Hanle curve. The dephasing time for this process is $T \approx \hbar\sqrt{N}/A$, where $A$=90µeV is the hyperfine constant in GaAs and $N$ is the number of nuclear spins within the wavefunction of the QD electron [27]. The measured spin lifetime of 16 ns corresponds to $5\times10^6$ nuclei or a diameter of 1500 nm (for 5 nm thickness). This is somewhat larger than a typical natural GaAs QD but within the expected size range. A more complete interpretation of the linewidth will require further study.

We pump the electron most efficiently when the laser intensity is high. The depth of the Hanle peak for $X^-$ is proportional to the degree of ground state electron polarization. We measure X⁻ polarization with and without a transverse magnetic field in order to see the peak depth as a function of bias and power. This is depicted in Fig. 2(c) by the thickness of the shaded regions for two laser intensities. At the higher laser intensity, polarized photogenerated electrons replace electrically-injected electrons, so the Hanle peak is relatively large (roughly 15%) for all biases. In contrast, for lower laser intensity the Hanle peak is only large near the charging threshold at 4V but disappears as the bias is increased. This change coincides with the increase in trion PL polarization and results from electrical injection of unpolarized electrons. These results show the clear correlation between the sign of X⁻ polarization and the degree of electron spin pumping (Hanle peak depth).

Measurements in a high longitudinal magnetic field complete the picture of optical orientation in single QDs. Each of the PL peaks splits into a Zeeman doublet [Figs 3(a) and 4(a)] and we measure doublet splittings and intensities obtained with both laser polarizations. A raw polarization $\rho_{raw} = (I_+ - I_-)/(I_+ + I_-)$ is obtained directly from the intensities $I_+$ and $I_-$ of the two peaks in the Zeeman doublet. $\rho_{raw}$ is shown in Fig. 3(b) as a function of the laser polarization, which is controlled by a variable retarder. The polarization memory is calculated using $\rho = (\rho_{raw}^+ - \rho_{raw}^-)/2$, where $\rho_{raw}^\pm$ are the raw polarizations obtained with the corresponding laser helicity $\sigma^\pm$. This removes the part of the raw polarization arising from thermalization between Zeeman levels. The amplitude of the curves in Fig. 3(b) corresponds to the polarization memory.

The high magnetic field restores the positive circular polarization of the neutral exciton (≈40%), which is otherwise suppressed by the anisotropic electron-hole exchange interaction in an asymmetric QD [29]. This occurs once the Zeeman splitting of the exciton becomes much larger than the exchange splitting, which is of order 10 μeV [29]. The high magnetic field changes the $X^+$ polarization very little. The $X^-$ polarization is also qualitatively similar at 5T, [Fig. 4(b)], and in particular, the negative polarization persists. This is important, because it implies that an alternative mechanism for negative polarization of $X^-$ that involves exchange interactions in triplet states [8] is less important here. Such a mechanism should turn off at high magnetic fields.

Finally, we demonstrate efficient optical pumping of the nuclear spins in all three charge states of the QD at 5T. When electrons are optically oriented, they can transfer polarization to the nuclear spins in the QD through a hyperfine flip-flop process [1,14]. Holes do not transfer their spins because they do not interact strongly with the nuclei. We measure nuclear polarization $P_N$ through the Overhauser effect, which changes the doublet splittings. Essentially, the nuclear polarization exerts an effective magnetic field on the electron spin, leading to an Overhauser shift [seen as the amplitudes of the curves in Fig. 3(c)]. The nuclear polarization tracks the electron polarization and can therefore be tuned with the applied bias [Fig. 4(c)]. When $X^o$ or $X^+$ is present in the QD, $P_N$ is large and mostly independent of bias. For $X^-$, $P_N$ starts out large at 4V bias, where most electrons are optically polarized, but decreases with increasing bias as unpolarized electrons are injected. This coincides with the increase in PL polarization and the suppression of the Hanle depolarization. For the highest laser pumping intensities, we have observed shifts of 81 μeV, corresponding to a degree of nuclear polarization $P_N = 60\%$. These shifts in the electron spin splitting would require an external magnetic field of 14T to achieve through the usual Zeeman interaction, and could be used to advantage as a way to suppress the influence of nuclear spin fluctuations [28] or as a form of long-lived quantum memory [30].

By combining the classic techniques of optical orientation with those of single dot spectroscopy, we reveal a much higher level of detail than that possible with ensemble measurements. We have observed dramatic differences in polarization behavior as we changed the charge state of a single dot from positive to neutral to negative, and we have demonstrated efficient optical pumping of electron and nuclear spin.

We acknowledge funding by DARPA/QuIST, ARDA/NSA/ARO, and CRDF.

Fig. 1

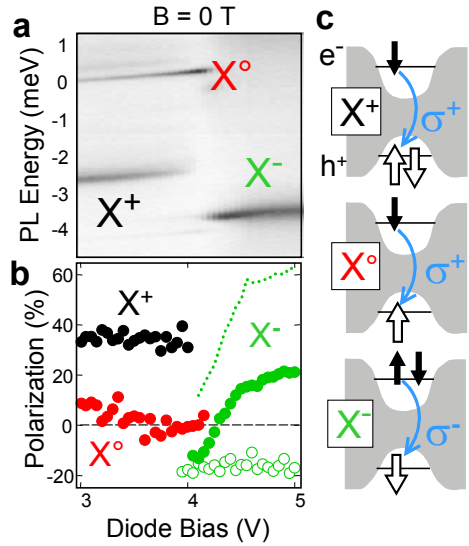

**Fig. 1.** (a) PL intensity (grayscale) for a single QD as a function of emitted photon energy and applied bias at zero magnetic field. Peaks are labeled for the neutral exciton (X°), negative trion (X⁻) and positive trion (X⁺). The PL energy scale is relative to the X° peak at 3V bias ($E_{X°}$ = 1.663 eV). The energy of the excitation laser was 1.691 eV ($E_{X°}$ + 28 meV). (b) PL polarization memory for the spectral lines in (a) (solid symbols). Open circles correspond to a higher excitation intensity. Dotted line corresponds to a lower excitation photon energy ($E_{X°}$ + 15 meV). (c) Schematic diagrams of QD band profiles and ground state spin configurations for all three charge states.

Fig. 2

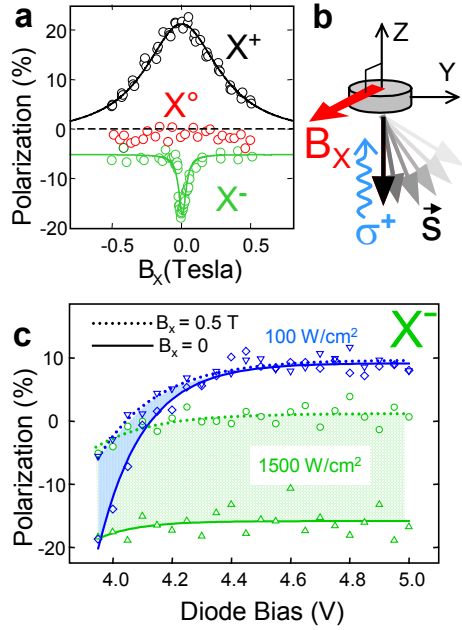

**Fig. 2.** **(a)** Hanle effect measurements for all three charge states. **(b)** Geometry for Hanle effect: electron spins are photogenerated with spins that are antiparallel to the propagation direction of the circularly polarized light (Z). Spins precess in the Y-Z plane, perpendicular to the transverse magnetic field direction (X). **(c)** Bias dependence of trion polarization for two different laser intensities and two transverse magnetic fields. The thickness of the shaded regions corresponds to the depth of the Hanle peak, which is related to the degree of ground state electron polarization.

Fig. 3

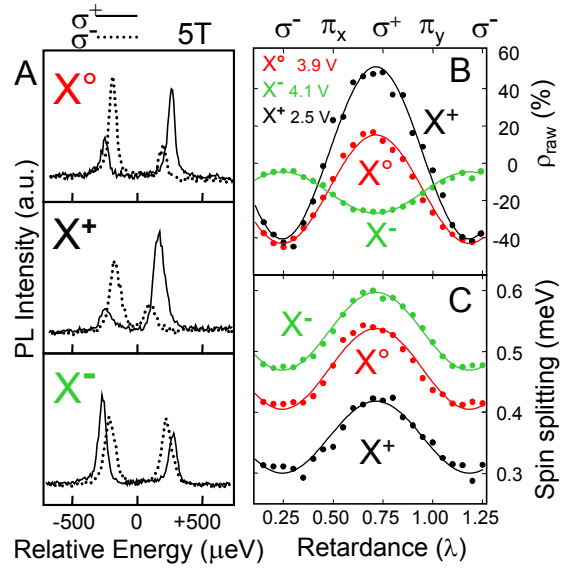

**Fig. 3.** (a) PL spectra of $X^o$, $X^-$ and $X^+$ in a longitudinal magnetic field (5T) for both polarizations of the laser. (b) Raw polarizations calculated from peak intensities in each Zeeman doublet, as a function of laser polarization. (c) Spin splitting (Zeeman splitting + Overhauser shift) as a function of laser polarization.

Fig. 4

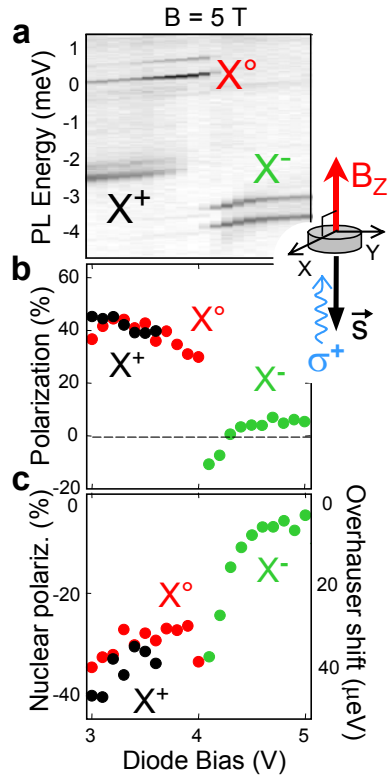

**Fig. 4. (a)** PL intensity (grayscale) for a single quantum dot as a function of the emitted photon energy and applied bias at 5T and $\sigma^-$ laser polarization. **(b)** Polarization memory calculated from peak intensities in (a) and in the analogous spectrum for $\sigma^+$ laser polarization. **(c)** Nuclear spin polarization (proportional to Overhauser shift) obtained from doublet splittings in (a) and in the analogous spectrum for $\sigma^+$ laser polarization.

**References**

[1] *Optical Orientation*, edited by F. Meier and B. Zakharchenya, *Modern Problems in Condensed Matter Sciences*, Vol. 8 (North-Holland, Amsterdam 1984).

[2] A. Imamoglu et al., Phys. Rev. Lett. **83**, 4204 (1999).

[3] C. Piermarocchi, P. Chen, L. J. Sham, D. G. Steel, Phys. Rev. Lett. **89**, 167402 (2002).

[4] F. Troiani, E. Molinari, U. Hohenester, Phys. Rev. Lett. **90**, 206802 (2003).

[5] T. Calarco et al., Phys. Rev. A **68**, 012310 (2003).

[6] A. Shabaev, A1. L. Efros, D. Gammon, I. A. Merkulov, Phys. Rev. B **68**, 201305 (2003).

[7] R. I. Dzhioev *et al.,* Phys. Solid State, **40** 1587 (1998).

[8] S. Cortez *et al.,* Phys. Rev. Lett. **89**, 207401 (2002).

[9] I. E. Kozin et al., Phys. Rev. B **65**, 241312 (2002).

[10] V. K. Kalevich et al., phys. stat. sol. (b) **238**, 250 (2003).

[11] R. J. Epstein et al., Appl. Phys. Lett. **78**, 733 (2001).

[12] K. Gündoğdu et al., Appl. Phys. Lett. 84, 2793 (2004).

[13] T. Flissikowski, I. A. Akimov, A. Hundt, F. Henneberger, Phys. Rev. B **68**, R161309 (2003).

[14] D. Gammon et al., Phys. Rev. Lett. **86**, 5176 (2001).

[15] J. G. Tischler, A. S. Bracker, D. Gammon, D. Park, Phys. Rev. B **66**, R081310 (2002).

[16] R. J. Warburton *et al.,* Nature **405**, 926 (2000).

[17] A. J. Shields, M. Pepper, M. Y. Simmons, D. A. Ritchie, Phys. Rev. B **52**, 7841 (1995).

[18] L. Landin et al., Science **280**, 262 (1998).

[19] A. Hartmann et al., Phys. Rev. Lett. **84**, 5648 (2000).

[20] D. V. Regelman et al., Phys. Rev B **64**, 165301 (2001).

[21] F. Findeis et al., Phys. Rev. B **63**, 121309 (2001).

[22] J. J. Finley et al., Phys. Rev. B **63**, 161305 (2001).

[23] M. Bayer et al., Phys. Rev. B **65**, 195315 (2002).

[24] T. C. Damen et al., Phys. Rev. Lett. **67**, 3432 (1991).

[25] G. Finkelstein *et al.,* Phys. Rev. B **58**, 12637 (1998).

[26] R. I. Dzhioev et al. Phys. Rev. B **66**, 153409 (2002).

[27] I. A. Merkulov, A1. L. Efros, M. Rosen, Phys. Rev. B **65**, 205309 (2002).

[28] A. V. Khaetskii, D. Loss, L. Glazman, Phys. Rev. Lett. **88**, 186802 (2002).

[29] D. Gammon et al., Phys Rev. Lett. **76**, 3005 (1996).

[30] J. M. Taylor, C. M. Marcus, M. D. Lukin, Phys. Rev. Lett. **90**, 206803 (2003).